\documentclass[]{pasj02} 
\usepackage[switch,mathlines]{lineno} 
\usepackage{natbib} 
\usepackage{graphicx}
\usepackage{amsmath}
\usepackage{bm}
\usepackage{color}

\jyear{2026}
\Received{}
\Accepted{}

\begin{document} 

\title{ Transient Signatures of Star-Envelope Collisions in Little Red Dots }

\author{
 Tomoya \textsc{Suzuguchi},\altaffilmark{1}\altemailmark\orcid{0009-0005-1459-1846} \email{suzuguchitm@ccs.tsukuba.ac.jp} 
 Kohei \textsc{Inayoshi},\altaffilmark{2}\orcid{0000-0001-9840-4959}
}
\altaffiltext{1}{Center for Computational Science, University of Tsukuba, 1-1-1 Tennodai, Tsukuba, Ibaraki 305-8577, Japan}
\altaffiltext{2}{Kavli Institute for Astronomy and Astrophysics, Peking University, Beiging 100871, China}




\maketitle

\begin{abstract}
Little red dots (LRDs) are compact high-redshift objects, newly discovered by the James Webb Space Telescope. Although LRDs exhibit broad Balmer emission lines suggestive of the presence of active galactic nuclei (AGN), their spectral features differ significantly from those of ordinary AGN. Recent studies suggest that their characteristics can be explained if accreting supermassive black holes (SMBHs) are embedded within dense gaseous envelopes and surrounded by compact stellar clusters. In this scenario, stars in the cluster can scatter onto plunging orbits that intersect the envelope and collide with its surface. Here we investigate the observational properties of such star-envelope collisions as luminous transient events. We find that collisions involving red supergiants with radii of $\sim 10^{3}~R_\odot$, together with gaseous envelopes whose masses are comparable to those of the central SMBHs, are the most promising targets due to their high luminosities and long durations. For compact clusters with sizes of $\lesssim 10~{\rm pc}$, such massive stars can participate in star–envelope collisions within their lifetimes at event rates reaching $\sim 0.3~{\rm yr}^{-1}$ per LRD. We show that these transients are detectable with future wide-field surveys such as the Nancy Grace Roman Space Telescope if they occur at relatively low redshifts ($z \lesssim 1$). Detection of such transients would provide strong evidence for the envelope+stellar-cluster scenario of LRDs and offer a unique probe of the envelope mass, which is otherwise difficult to constrain from LRD spectra alone. 
\end{abstract}


\section{Introduction}
Little red dots (LRDs) are compact, red objects newly discovered by the \textit{James Webb Space Telescope} (JWST)~\citep[e.g.,][]{Kocevski+2023,Matthee+2024,Greene+2024}. A large fraction of photometrically selected LRDs exhibit broad Balmer emission lines~\citep[e.g.,][]{Hviding+2025}, suggesting the presence of active galactic nuclei (AGN). Their defining feature is a ``V-shaped'' spectral energy distribution (SED) extending from the rest-frame ultraviolet (UV) to optical wavelengths, characterized by a blue UV continuum and a significantly redder optical component~\citep[e.g.,][]{Kocevski+2025,Setton+2024}. Additional observational properties, including X-ray and radio faintness~\citep[e.g.,][]{Maiolino+2025,Akins+2025,Mazzolari+2026,Gloudemans+2025}, weak hot dust emission~\citep[e.g.,][]{Williams+2024,Perez-Gonzalez+2024,Setton+2025,Akins+2025}, and little or no detectable variability~\citep[e.g.,][]{Kokubo&Harikane2025,Tee+2025,Liu+2026}, further distinguish LRDs from typical AGN. While some studies have suggested that at least part of the observed properties may be explained by compact stellar clusters~\citep[e.g.,][]{Zhang+2026,Zwick+2026,Liempi+2026,Chisholm+2026}, the physical nature of LRDs remains under active debate. 

Recent observational and theoretical studies suggest that the enigmatic multi-wavelength properties of LRDs can be explained by supermassive black holes (SMBHs) embedded in dense gaseous environments~\citep[e.g.,][]{Juodvbalis+2024,Inayoshi&Maiolino2025,Ji+2025}. Several LRDs exhibit a pronounced Balmer break near the inflection point of the V-shaped continuum spectrum (rest-frame $\simeq 3600$~\AA), which is difficult to reproduce with stellar emission alone~\citep[e.g.,][]{Naidu+2025,de_Graaff+2025a,Taylor+2025}. In addition, absorption features on top of the broad Balmer lines (particularly evident in high-resolution spectroscopy) indicate the presence of dense gas absorbers surrounding the nuclear region~\citep[e.g.,][]{Lin+2024,D_Eugenio+2025,Matthee+2026}. The red optical continuum can be reproduced by blackbody emission from an optically thick envelope with an effective temperature of $\sim 5000~{\rm K}$~\citep[e.g.,][]{Kido+2025,Liu+2025,Naidu+2025,de_Graaff+2025a,Inayoshi+2026b,Umeda+2026}. In this scenario, the SMBH is surrounded by a Compton-thick medium that obscures X-rays from the accretion disk corona~(e.g., \citealt{Juodvbalis+2024,Maiolino+2025,Rusakov+2026}; see also \citealt{Pacucci&Narayan2024,Madau&Haardt2024,Inayoshi+2025}). If such envelopes are indeed present, the interaction between the envelope and relativistic jets launched from the accretion disk may drive efficient hadronic processes, leading to copious production of high-energy cosmic rays and neutrinos~\citep[e.g.,][]{Kuze+2026,Yoshida&Meier2026}.  

Despite the success of the AGN+dense gas model in explaining many observed properties of LRDs, the origin of the relatively flat UV component in the V-shaped SED remains uncertain. One possible explanation is stellar emission from a dense, compact ($\sim 10~{\rm pc}$) stellar cluster surrounding the SMBH~\citep[e.g.,][]{Inayoshi+2026a,Asada+2026} and in some cases, their host galaxies~\citep[e.g.,][]{Killi+2024,Golubchik+2025,Sun+2026}. In this picture, LRDs can be understood as an ``envelope + dense stellar cluster" configuration. Such a compact stellar environment may trigger a variety of high-energy astrophysical phenomena. In particular, the dense stellar cluster surrounding the envelope is expected to enhance the rate of tidal disruption events (TDEs)~\citep[e.g.,][]{Magorrian&Tremaine1999,Wang&Merritt2004,Stone&Metzger2016,Chang+2025}. TDEs occur when stars are scattered onto orbits that bring them sufficiently close to the SMBH to be disrupted by tidal forces~(\citealt{Rees1988,Phinney1989,Gezari2021} and references therein). Since a more compact stellar distribution increases the probability of such close encounters, the TDE rate is substantially boosted~\citep[e.g.,][]{Wang&Merritt2004}. In particular, the compact nuclear structure inferred for LRDs may provide an ideal environment for frequent TDE activity~(\citealt{Inayoshi+2024,Bellovary2025}; see also \citealt{Kar_Chowdhury+2024}). These high-energy phenomena not only provide probes of the LRD scenario but also offer insights into the stellar population in the surrounding compact clusters. 

In this paper, we investigate interactions between stars and the envelope prior to tidal disruption. In dense stellar clusters, stars can be scattered onto plunging orbits that intersect the envelope before reaching the tidal disruption radius of the SMBH. Such encounters lead to ``star-envelope collisions", potentially producing transient flares that probe the physical properties of the envelope. This mechanism is analogous to quasi-periodic eruptions (QPEs), recurring X-ray flares observed in galactic nuclei~\citep[e.g.,][]{Miniutti+2019,Giustini+2020,Arcodia+2021,Arcodia+2024a,Arcodia+2025,Chakraborty+2021,Quintin+2023,Nicholl+2024,Hernandez-Garcia+2025,Chakraborty+2025a,Bykov+2025,Baldini+2026}, which are thought to originate from repeated star-disk interactions~\citep[e.g.,][]{Linial&Metzger2023,Franchini+2023,Tagawa&Haiman2023,Zhou+2024a,Zhou+2024b,Yao+2025,Vurm+2025,Huang+2025,Suzuguchi&Matsumoto2026}. Motivated by this analogy, we adapt the theoretical framework developed for QPEs to predict the observational signatures of star-envelope collisions in LRDs.

This paper is organized as follows. In Section~\ref{sec:collision}, we present the formalism for calculating the observational signatures of star-envelope collisions and describe the parameter space considered in this work. We then present the resulting observables, including spectral properties, and discuss the parameter ranges relevant for detection. In Section~\ref{sec:event_rate}, we estimate the event rate of the star-envelope collision based on TDE rate calculations. In Section~\ref{sec:observability}, we assess their observability with current and future facilities. Finally, in Section~\ref{sec:discussion}, we discuss several implications, including the expected number of observable events and the connection between envelope properties and TDE observability.

\section{Stellar Collisions with Envelopes} \label{sec:collision}
In models for the V-shaped SEDs of LRDs, the red optical continuum is attributed to blackbody emission from an optically thick envelope, while the blue UV emission originates from young stars in a dense stellar cluster surrounding the SMBH+envelope system~\citep[e.g.,][]{Naidu+2025,Liu+2025,Kido+2025,de_Graaff+2025b,Inayoshi+2026a,Umeda+2026,Asada+2026}. In such a configuration, stars in the dense cluster can be scattered onto orbits that intersect the envelope. These interactions may produce transients through collisions between stars and the envelope surface. 

When a star enters the gaseous envelope at supersonic velocity, it drives a shock that compresses and heats the surrounding gas. The shocked material is initially optically thick and subsequently expands under its own high pressure. As the ejecta expands, its optical depth decreases and radiation begins to escape once the photon diffusion timescale becomes comparable to the dynamical expansion timescale. 

\begin{figure}[t]
  \begin{center}
  \includegraphics[keepaspectratio, scale=0.4]{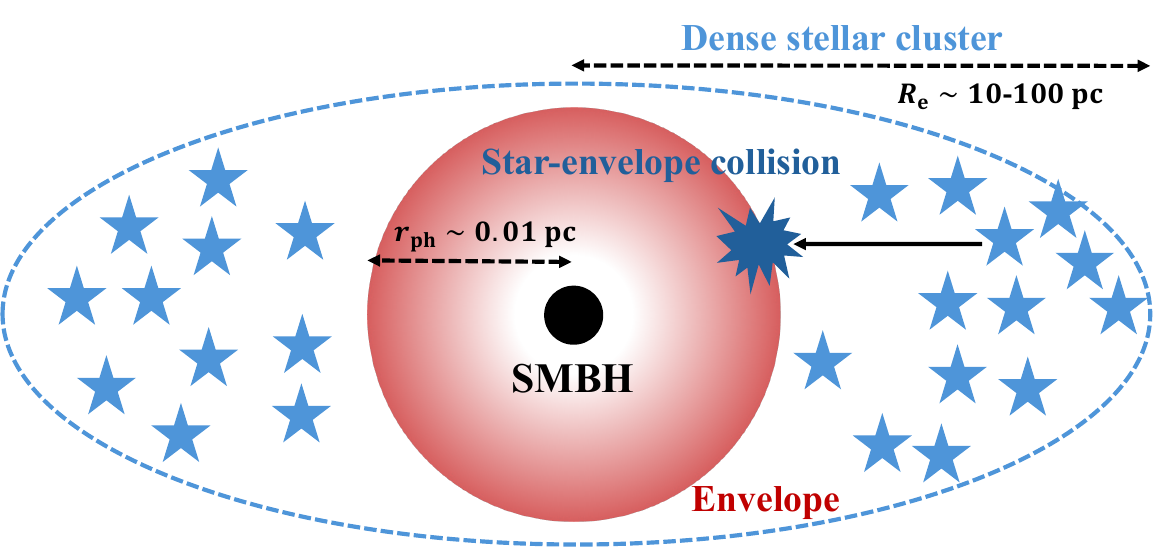}
  \end{center}
  \caption{A schematic view of stellar collisions with gaseous envelopes in LRDs. Thermal emission from the envelope produces the red optical continuum of the V-shaped SEDs, while stellar emission from the surrounding dense cluster accounts for the blue UV emission. When stars in the cluster fall onto the envelope surface, supersonic star-envelope collisions produce transient flares. }
  \label{fig:schematic}
\end{figure}

In this section, we calculate the observational signatures of star-envelope collisions, including their durations, peak luminosities, and peak temperatures. We also discuss the expected spectral properties of these transients. A schematic illustration of the scenario is shown in Figure~\ref{fig:schematic}. 

\subsection{Structure of Envelopes} \label{subsec:structure}
Modeling star-envelope collisions requires specifying the structure of the gaseous envelope around an accreting SMBH. For simplicity, we assume that the SMBH is fully enshrouded by a smooth, spherically symmetric envelope. Such a configuration has been proposed to explain several observed properties of LRDs, including their red optical continua, X-ray and radio faintness, and Balmer breaks and absorption features~\citep[e.g.,][]{Inayoshi&Maiolino2025,Naidu+2025,Liu+2025,Kido+2025,de_Graaff+2025b,Inayoshi+2026a,Umeda+2026}. In reality, however, the envelope structure may deviate from spherical symmetry if the accreting gas possesses substantial angular momentum. Recent observations suggest that the obscuring medium may be clumpy rather than fully covering the SMBH~\citep[e.g.,][]{Ji+2026,Tang+2026}. Given these uncertainties, we adopt the simplest idealized model in which the envelope is spherical and completely surrounds the accreting SMBH. 

The emission from the envelope is assumed to originate from blackbody radiation at the photosphere, which can reproduce the red optical continuum emission in LRDs~\citep{Kido+2025,Umeda+2026}. The photospheric radius, corresponding to the characteristic size of the envelope, is determined by the luminosity $L_{\rm ph}$ and effective temperature $T_{\rm ph}$ as
\begin{align}
    \label{eq:envelope_size}
    r_{\rm ph} \simeq \left(\frac{L_{\rm ph}}{4 \pi \sigma_{\rm SB} T_{\rm ph}^{4}}\right)^{1/2}
    \simeq 6.4 \times 10^{15}~{\rm cm}~ 
    \frac{L_{\rm ph,10}^{1/2}}{T_{\rm ph,0}^{2}},
\end{align}
where $\sigma_{\rm SB}$ denotes the Stefan-Boltzmann constant, and $L_{\rm ph} = 10^{10}~L_{\odot}~L_{\rm ph,10}$ and $T_{\rm ph} = 6000~{\rm K}~T_{\rm ph,0}$ are fiducial values in the following discussion~\citep[e.g.,][]{Inayoshi&Ho2025}.

The next step is to specify the radial density profile of the envelope. This issue has been extensively studied in the context of optical/UV emission from TDEs. Although the origin of optical/UV emission in TDEs remains uncertain, one promising scenario attributes it to thermal emission from a gaseous envelope produced by disrupted stellar debris~\citep[e.g.,][]{Loeb&Ulmer1997,Coughlin&Begelman2014,Metzger2022}. Recent numerical simulations tracking the evolution of TDEs from stellar disruption to fallback accretion suggest that an optically thick envelope naturally forms around the SMBH~\citep[e.g.,][]{Ryu+2023b,Price+2024,Steinberg&Stone2024}. 

In this work, we adopt the envelope model of the pioneering study by \citet{Loeb&Ulmer1997}, in which radiation pressure supports the envelope against the SMBH gravity. In this model, the density profile follows $\rho \propto r^{-\zeta}$ with a power-law index $\zeta = 3$, corresponding to an $n = 3$ polytropic equation of state. 
The density normalization is set by the total envelope mass of $M_{\rm env} = 10^{6}~M_{\odot}~M_{\rm env,6}$, whose choice is discussed in Section~\ref{subsec:parameter_spectrum}, and thus the density profile is given as
\begin{align}
    \label{eq:envelope_density}
    \rho_{\rm env}(r) &\simeq \frac{M_{\rm env}}{4 \pi r_{\rm ph}^{3} \ln{\xi}} 
    \left(\frac{r}{r_{\rm ph}}\right)^{-3}  \\
    &\simeq 5.9 \times 10^{-10}~{\rm g}~{\rm cm}^{-3}~
    \frac{M_{\rm env,6} T_{\rm ph,0}^{6}}{L_{\rm ph,10}^{3/2} \ln{\xi}}
    \left(\frac{r}{r_{\rm ph}}\right)^{-3},\nonumber
\end{align}
where $\xi \equiv r_{\rm ph}/r_{\rm min}$ and $r_{\rm min}$ is the inner boundary radius of the envelope. 

To characterize the thermodynamic state of the envelope and the strength of shocks generated by star-envelope collisions, we estimate the sound speed within the envelope. Under the assumption of hydrostatic equilibrium, one obtains
${\rm d}p/{\rm d}r = GM_{\bullet+{\rm env}}\rho_{\rm env}/r^{2}$, where $p$ is the pressure and $M_{\bullet+{\rm env}}$ is the total mass of the SMBH and envelope. For the density profile $\rho \propto r^{-3}$, the pressure gradient can be written as ${\rm d}p/{\rm d}r = 3 c_{\rm s}^{2} \rho / r$, where the sound speed is defined by $c_{\rm s} = ({\rm d}p/{\rm d}\rho)^{1/2}$. Substituting this relation into the hydrostatic equilibrium equation yields
\begin{align}
    \label{eq:sound_velocity}
    c_{\rm s}(r) &= \left(\frac{G M_{\bullet+{\rm env}}}{3 r}\right)^{1/2} \nonumber \\
    &\simeq 8.3 \times 10^{2}~{\rm km}~{\rm s}^{-1}~
    \frac{M_{\bullet + {\rm env},6}^{1/2} T_{\rm ph,0}}{L_{\rm ph,10}^{1/4}}
    \left(\frac{r}{r_{\rm ph}}\right)^{-1/2},
\end{align}
where $M_{\bullet+{\rm env}} = 10^{6}~M_{\odot}~M_{\bullet+{\rm env},6}$.

\subsection{Model of Stellar Collisions} \label{subsec:model}
When a star from the surrounding stellar cluster enters the envelope, it collides with the photosphere at a velocity
\begin{align}
    \label{eq:stellar_velocity}
    v_{\star} &\simeq \left(\frac{2 G M_{\bullet+{\rm env}}}{r_{\rm ph}}\right)^{1/2} \nonumber \\
    &\simeq 2.0 \times 10^{3}~{\rm km}~{\rm s}^{-1}~ 
    \frac{M_{\bullet+{\rm env},6}^{1/2} T_{\rm ph,0}}{L_{\rm ph,10}^{1/4}},
\end{align}
where $G$ is the gravitational constant. This velocity exceeds the sound speed at the envelope surface (Equation~\eqref{eq:sound_velocity}), implying that the stellar motion is supersonic. As a result, a bow shock forms ahead of the star and compresses the surrounding gas near the envelope surface. The shocked gas is heated and expands outward due to its high pressure with a characteristic velocity comparable to the stellar velocity, $v_{\rm ej} \sim v_{\star}$. After the star penetrates to a depth $d_{\star} = v_{\star} t$, where $t$ is the elapsed time since the collision begins, the ejecta mass becomes comparable to the mass swept up by the bow shock, 
\begin{align}
    \label{eq:ejecta_mass}
    M_{\rm ej} \simeq \pi \rho_{\rm env}(r_{\rm ph}) R_{\star}^{2} d_{\star},
\end{align}
where $R_{\star} = 10^{0} \, R_{\odot} \, R_{\star,0}$ is the stellar radius. Here we neglect the radial variation of the envelope density and evaluate $\rho_{\rm env}$ at the photosphere, since our analysis focuses on the early phase immediately after the star enters the envelope (see Equation~\eqref{eq:critical_depth} below). 

As the shocked material expands, the ejecta radius evolves as $R_{\rm ej} \sim v_{\rm ej} t$. Initially, the ejecta is opaque because of its high density and compact size. The optical depth is approximately $\tau(t) \sim \kappa M_{\rm ej}/(\pi R_{\rm ej}^{2}) \propto t^{-2}$, where $\kappa$ is the Thomson opacity. As the ejecta expands, its optical depth decreases and photons begin to escape once the diffusion timescale becomes comparable to the dynamical timescale, corresponding to the condition of $\tau(t) \sim c / v_{\rm ej}$, where $c$ is the speed of light. The corresponding diffusion timescale is
\begin{align}
    \label{eq:diffusion_time}
    t_{\rm diff} \simeq \left(\frac{\kappa M_{\rm ej}}{4 \pi c v_{\rm ej}}\right)^{1/2},
\end{align}
which characterizes the rise time, decay time, and overall duration of the transient to within a factor of a few~\citep[e.g.,][]{Arnett1980}.

In order for transients powered by star-envelope collisions to be directly observable, the ejecta must break out from the envelope before most photons escape. Otherwise, diffusive photons remain embedded within the optically thick envelope and may be absorbed or reprocessed by the surrounding gas. This condition is determined by comparing the diffusion timescale with the ejecta crossing time through the envelope, $t_{\rm cross} \simeq d_{\star}/v_{\rm ej}$. The diffusion timescale depends on the penetration depth $d_{\star}$ through the ejecta mass $M_{\rm ej} \propto d_{\star}$ (Equation~\eqref{eq:diffusion_time}), yielding $t_{\rm diff} \propto d_{\star}^{1/2}$, while the crossing timescale scales as $t_{\rm cross} \propto d_{\star}$. The critical depth is obtained by setting $t_{\rm diff}=t_{\rm cross}$ as
\begin{align}
    \label{eq:critical_depth}
    d_{\rm crit} &:= \frac{\kappa v_{\star} R_{\star}^{2} \rho_{\rm env}(r_{\rm ph})}{4 \pi c}, \nonumber \\
    &\simeq 1.0 \times 10^{8}~{\rm cm}~
    \left(\frac{\kappa}{0.4~{\rm cm}^{2}~{\rm g}^{-1}}\right)
    \left(\frac{R_{\star}}{R_{\odot}}\right)^{2} \nonumber \\
    &\quad \times \left(\frac{v_{\star}}{2 \times 10^{3}~{\rm km}~{\rm s}^{-1}}\right)
    \left(\frac{\rho_{\rm env}(r_{\rm ph})}{10^{-10}~{\rm g}~{\rm cm}^{-3}}\right).
\end{align}
For $d_{\star} < d_{\rm crit}$, most photons escape after the ejecta breaks out of the envelope surface, so that the diffusion-powered emission can be observed. In contrast, for $d_{\star} > d_{\rm crit}$, photons diffuse out while the ejecta is still embedded in the envelope. In the latter case, the escaping radiation may undergo substantial reprocessing by the surrounding gas, potentially modifying the observable signal and reducing the peak luminosity. In our analysis, we neglect these reprocessing effects in spectral modeling and focus on the shallow penetration regime, $d_{\star} \lesssim d_{\rm crit}$. Thus, the maximum duration time of the transient is computed at $d_{\star} = d_{\rm crit}$ as 
\begin{align}
    \label{eq:duration}
    t_{\rm dur} &= t_{\rm diff}(d_{\star}= d_{\rm crit}) \nonumber \\
    &\simeq 5.4~{\rm s}~
    \frac{R_{\star,0}^{2} \kappa_{\rm es,0} T_{\rm ph,0}^{6} M_{\rm env,6}}{(\ln{\xi}) L_{\rm ph,10}^{3/2}},
\end{align}
where $\kappa_{\rm es} = 0.4~{\rm cm}^{2}~{\rm g}^{-1}~\kappa_{\rm es,0}$. Note that the duration is independent of the SMBH mass. Although the SMBH mass enters the duration only through the stellar velocity (Equation~\eqref{eq:stellar_velocity}), this dependence cancels out because two competing effects offset each other: a higher stellar velocity causes faster ejecta expansion, shortening the diffusion time, while simultaneously allowing the star to penetrate deeper into the envelope, thereby increasing the ejecta mass and prolonging the diffusion time. 

Assuming equipartition between the kinetic and internal energy of the ejecta, the internal energy of the shocked material with a mass $M_{\rm ej,0}$ at $d_\star = d_{\rm crit}$ is 
\begin{align}
    E_{\rm ej} \simeq M_{\rm ej,0} v_{\star}^{2}.
\end{align}
Since the ejecta is initially opaque, its expansion is approximately adiabatic, implying $E \propto V^{-1/3}$, where $V$ is the ejecta volume. The internal energy at the diffusion time ($t = t_{\rm diff}$) is therefore 
\begin{align}
    E_{\rm diff} \simeq \left(\frac{V_{\rm diff}}{V_{\rm ej}}\right)^{-1/3} E_{\rm ej},
\end{align}
where $V_{\rm ej} \sim (1/7)\pi R_{\star}^{2} d_{\rm crit}$ is the volume of the shocked material at $d_{\rm crit}$ and $V_{\rm diff} \sim 4 \pi (v_{\rm ej} t_{\rm dur})^{3}$ is the ejecta volume at the diffusion time. The peak luminosity is then obtained as 
\begin{align}
    \label{eq:luminosity}
    L_{\rm peak} &\simeq \frac{E_{\rm diff}}{t_{\rm dur}} \\ \nonumber
    &\simeq 1.8 \times 10^{38}~{\rm erg}~{\rm s}^{-1}~
    \frac{R_{\star,0}^{4/3} T_{\rm ph,0}^{13/3} M_{\bullet+{\rm env},6}^{7/6} M_{\rm env,6}^{1/3}}{(\ln{\xi})^{1/3} \kappa_{\rm es,0}^{1/6} L_{\rm ph,10}^{13/12} }.
\end{align}

Finally, the photospheric temperature at the peak luminosity is evaluated by assuming that the ejecta emission is approximated by blackbody radiation\footnote{Equation~\eqref{eq:temperature} provides a lower limit on the temperature. If the shocked medium is photon-starved and thermal equilibrium is not fully established, the temperature depends on the efficiency of photon production~\citep[e.g.,][]{Nakar&Sari2010}. Less efficient photon production results in higher temperatures, since the available energy is distributed among fewer photons. For simplicity, we neglect these effects in this work.},
\begin{align}
    \label{eq:temperature}
    k_{\rm B}T_{\rm peak} &\simeq k_{\rm B}
    \left(\frac{L_{\rm peak}}{4\pi \sigma_{\rm SB} R_{\rm diff}^{2} }\right)^{1/4} \nonumber \\
    & \simeq 59~{\rm eV}~ 
    \frac{L_{\rm ph,10}^{2/3}}{R_{\star,0}^{2/3} \kappa_{\rm es,0}^{2/3} M_{\rm env,6}^{5/12} M_{\bullet+{\rm env},6}^{1/12} T_{\rm ph,0}^{8/3}},
\end{align}
where $R_{\rm diff} \simeq v_{\rm ej} t_{\rm dur}$ is the ejecta radius at the diffusion time.

\subsection{Parameter Dependencies \& Spectrum} \label{subsec:parameter_spectrum}
Thus far, the model is characterized by six parameters: the photospheric luminosity $L_{\rm ph}$, photospheric temperature $T_{\rm ph}$, the radial extent of the envelope $\xi = r_{\rm ph}/r_{\rm min}$, SMBH mass $M_{\bullet}$, envelope mass $M_{\rm env}$, and the radius of the injected star $R_{\star}$. In the following, we discuss plausible ranges for these parameters and relate them to observed properties of LRDs.

The photospheric luminosity corresponds to the rest-frame optical luminosity of observed LRDs, typically $10^{10-11} \, L_{\odot}$. We thus consider two representative cases, $L_{\rm ph} = 10^{10}~L_{\odot}$ and $10^{11}~L_{\odot}$. Observations suggest that the photospheric temperature of LRDs lies within a relatively narrow range of $4000$-$7000~{\rm K}$~\citep[e.g.,][]{Umeda+2026}. We therefore adopt a fiducial value of $T_{\rm ph} = 6000~{\rm K}$. The parameter $\ln{\xi}$ remains uncertain because the inner radius of the envelope $r_{\rm min}$ is poorly constrained both observationally and theoretically. For simplicity, we adopt $\ln{\xi} = 1.0$ throughout this work.

The SMBH mass is related to the photospheric luminosity through the Eddington ratio, $\lambda_{\rm Edd} = L_{\rm ph} / L_{\rm Edd}$, yielding
\begin{align}
    \label{eq:SMBH_mass}
    M_{\bullet} &\simeq 3.1 \times 10^{5}~M_{\odot}~
    \lambda_{\rm Edd}^{-1} 
    L_{\rm ph,10},
\end{align}
where $L_{\rm Edd} = 4 \pi c G M_{\bullet}/\kappa_{\rm es}$ is the Eddington luminosity. The Eddington ratio is not directly constrained observationally. However, from a theoretical perspective, \citet{Loeb&Ulmer1997} argued that radiation-pressure-supported envelopes naturally maintain luminosities close to the Eddington limit over a wide radial range. Motivated by this expectation, we adopt two representative values, $\lambda_{\rm Edd} = 1.0$ and $0.5$. 

Determining the envelope mass is difficult both observationally and theoretically. Nevertheless, its maximum value can be constrained by requiring that the radiative output does not exceed the Eddington luminosity~\citep{Kido+2025}. This condition suggests that the envelope mass is unlikely to exceed the SMBH mass\footnote{If LRDs are associated with quasi-star, however, the envelope mass may exceed the SMBH mass~\citep[e.g.,][]{Begelman&Dexter2026}.}. Thus, we treat the envelope mass as a free parameter with an upper limit of $M_{\rm env} \lesssim M_{\bullet}$. 

The stellar radius of stars colliding with the envelope is also uncertain. In general, most stars scattered toward the SMBH are expected to be predominantly solar-type main-sequence stars with typical radii of $\simeq R_{\odot}$, since the stellar mass function peaks near solar mass and main-sequence lifetimes are much longer than those of post-main-sequence phases. However, a fraction of stars can be in more evolved stages, where stellar radii can reach $\sim 10$-$100~R_\odot$. In some rare cases involving evolved massive stars, the radius may exceed $10^{3}~R_\odot$. To capture this range, we explore cases with $1\leq R_{\star}/R_{\odot}\leq 10^{3}$. 

\begin{figure}[t]
  \begin{center}
  \includegraphics[keepaspectratio, scale=0.38]{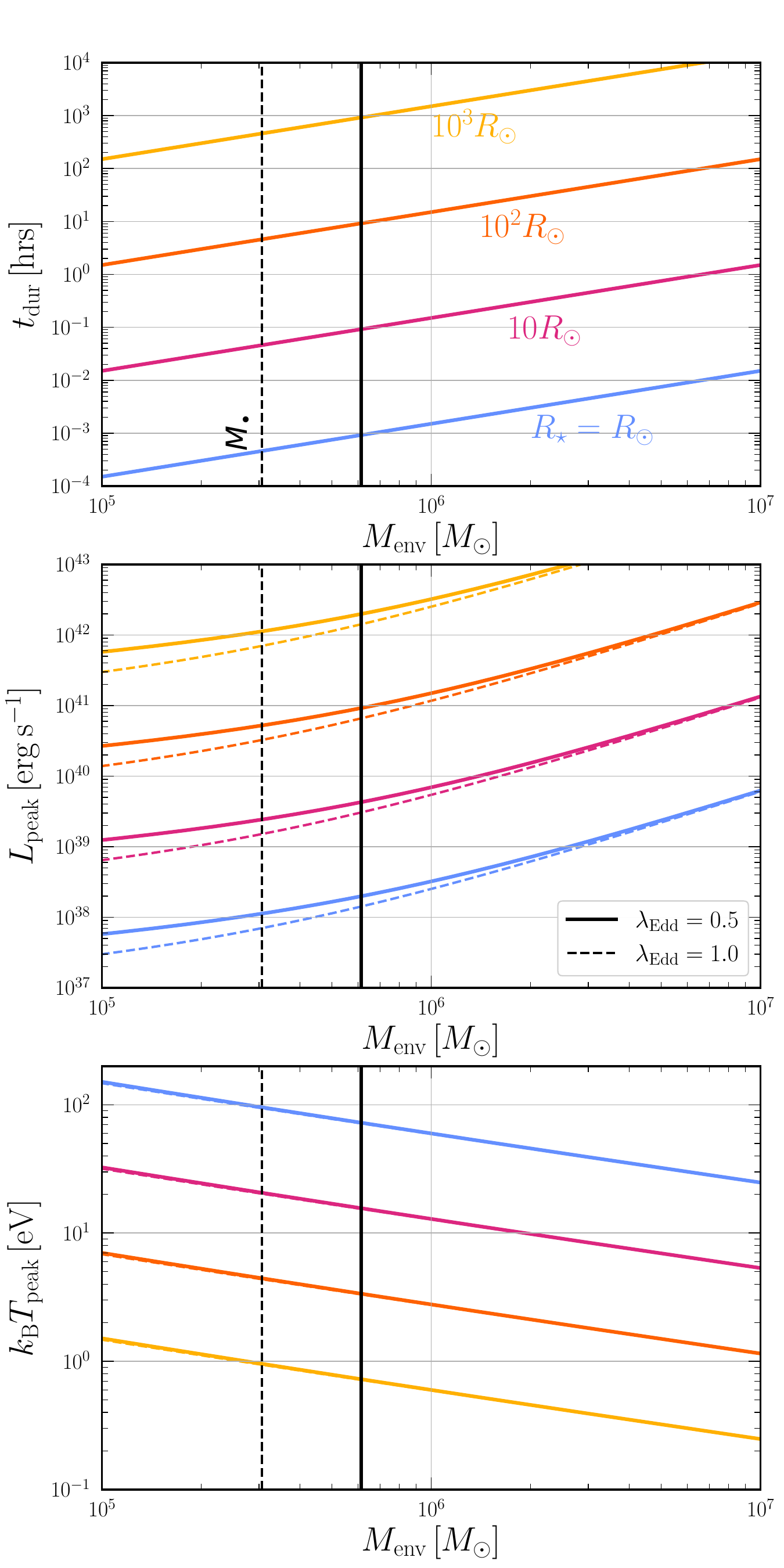}
  \end{center}
  \caption{Observable quantities of star-envelope collisions as functions of envelope mass, including durations (\textit{upper panel}), peak luminosities (\textit{middle panel}), and peak temperatures (\textit{lower panel}). The photospheric luminosity of the LRD envelope is set to $L_{\rm ph} = 10^{10}~L_{\odot}$. Different colors show different stellar radii: $R_{\star} = R_{\odot}$ (blue), $10~R_{\odot}$ (magenta), $10^{2}~R_{\odot}$ (orange), and $10^{3}~R_{\odot}$ (yellow). The solid and dashed lines correspond to Eddington ratios of $\lambda_{\rm Edd} = 0.5$ and $1.0$, respectively. The vertical lines represent the envelope masses corresponding to $M_{\rm env} = M_{\bullet}$ for each Eddington ratio. }
  \label{fig:obs_lowlum}
\end{figure}

Figure~\ref{fig:obs_lowlum} summarizes the observable properties of star-envelope collisions; the duration, peak luminosity, and peak temperature (from the top to the bottom) for the case of $L_{\rm ph} = 10^{10}~L_{\odot}$. The corresponding results for $L_{\rm ph} = 10^{11}~L_{\odot}$ are qualitatively similar and are shown in Appendix~\ref{appsec:parameter}. The horizontal axis represents the envelope mass with different colors indicating different stellar radii. The duration and peak luminosity increase with increasing envelope mass, while the peak temperature decreases. These general trends can be understood as follows. As the envelope mass increases, the density at the collision site increases, leading to a larger ejecta mass. This in turn increases both the diffusion timescale and the total radiated energy, resulting in longer durations and higher peak luminosities. At the same time, the longer diffusion timescale allows the ejecta to expand further before photons escape, which reduces the blackbody temperature.

Figure~\ref{fig:obs_lowlum} also shows that the larger stellar radii lead to longer durations and higher luminosities, while the temperature decreases. The physical reason for these trends is similar to that obtained when increasing the envelope mass. In both cases, a larger interaction scale (i.e., stellar size) produces a more massive ejecta, which increases the diffusion timescale and the total radiation energy. As a result, the emission becomes longer-lived and more luminous, while the photosphere expands to larger radii, lowering the effective temperature. Overall, collisions involving larger-size stars and more massive envelopes are more favorable for observational detection. In the remainder of this section, as well as in Sections~\ref{sec:event_rate} and \ref{sec:observability}, we adopt the most optimistic case, $M_{\rm env} = M_{\bullet}$ and $R_{\star} = 10^{3}~R_{\odot}$. 

\begin{figure}[t]
  \begin{center}
  \includegraphics[keepaspectratio, scale=0.42]{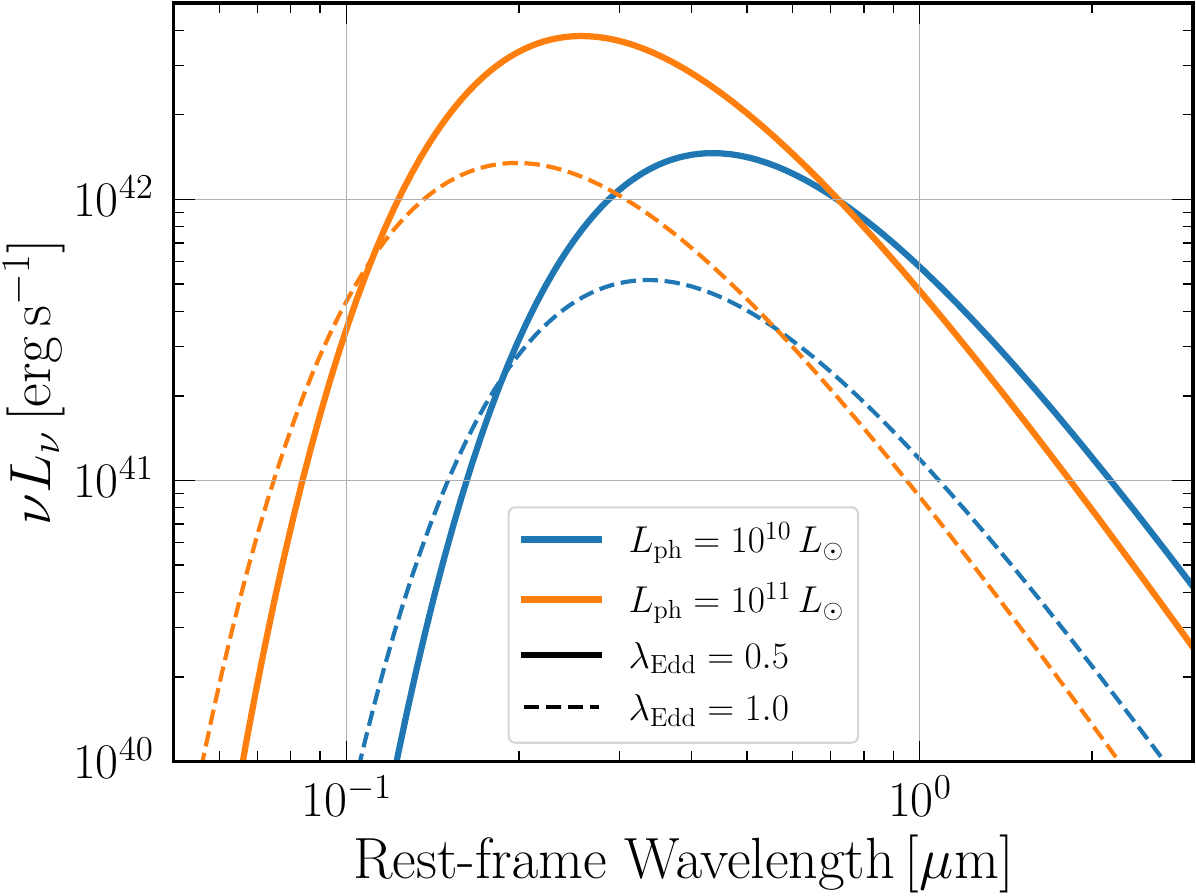}
  \end{center}
  \caption{Spectra of star-envelope collisions. The horizontal axis shows the rest-frame wavelength. Different colors represent different photospheric luminosities: $L_{\rm ph} = 10^{10}~L_{\odot}$ (blue) and $L_{\rm ph} = 10^{11}~L_{\odot}$ (orange). The solid and dashed lines correspond to Eddington ratios of $\lambda_{\rm Edd} = 0.5$ and $1.0$, respectively. The stellar radius injected into the envelope and the envelope mass are fixed at $R_{\star} = 10^{3}~R_{\odot}$ and $M_{\rm env} = M_{\bullet}$, respectively. }
  \label{fig:SED}
\end{figure}

Figure~\ref{fig:SED} presents the radiation spectra of star-envelope collision transients at their peak luminosities for two representative LRD luminosities, $L_{\rm ph} = 10^{10}~L_{\odot}$ (blue) and $10^{11}~L_{\odot}$ (orange) with two cases of Eddington ratios $\lambda_{\rm Edd} =0.5$ (solid) and $1.0$ (dashed). As discussed above, we adopt the fiducial parameters, $M_{\rm env} = M_{\bullet}$, $T_{\rm ph} = 6000~{\rm K}$, and $R_{\star} = 10^{3}~R_{\odot}$. The horizontal axis represents the rest-frame wavelength. Note that the underlying LRD spectrum itself is not included in this figure. The transient emission is approximated as blackbody radiation,
\begin{align}
    \nu L_{\nu} &= 4 \pi^{2} R_{\rm diff}^{2} \nu B_{\nu}(T_{\rm BB}),
\end{align}
where $B_{\nu}(T)$ is the Planck function. The transient luminosity reaches $\sim 10^{42}~{\rm erg}~{\rm s}^{-1}$ with spectral peaks at wavelengths of $0.2$-$0.5~\mu{\rm m}$ in the rest-frame UV band. Given that the UV luminosity of LRDs is typically $\sim 10^{43}~{\rm erg}~{\rm s}^{-1}$, star-envelope collisions could produce UV variability at the level of $\sim 10 \%$. 

Variations in the photospheric luminosity, or equivalently the SMBH mass, have only a weak impact on the transient properties in the case of $M_{\rm env} = M_{\bullet}$. This result follows from Equations~\eqref{eq:duration} and \eqref{eq:luminosity}, which give relatively weak scalings, $t_{\rm dur} \propto M_{\bullet}^{-1/2}$ and $L \propto M_{\bullet}^{5/12}$. As a result, the characteristic duration and luminosity of the transients depend only weakly on $L_{\rm ph}$ (or $M_\bullet$), implying that star-envelope collisions may be observable over a broad range of LRD luminosities.

\section{Event Rates} \label{sec:event_rate}
In this section, we estimate the event rate of star-envelope collisions. If LRDs are surrounded by dense stellar clusters, two-body relaxation can continuously scatter stars onto low-angular-momentum orbits that intersect the envelope surface. 

To evaluate the collision rate, we first estimate the properties of the surrounding stellar clusters.
The stellar mass of the cluster ${\cal M}_{\star}$ is inferred from the UV luminosity through
\begin{align}
    {\cal M}_{\star} \simeq 2.3 \times 10^{8}~M_{\odot}~
    \left(\frac{L_{\rm UV}}{10^{9} L_{\odot}}\right)
    \left(\frac{t_{\rm age}}{10^{9} \, {\rm yr}}\right),
\end{align}
where $t_{\rm age}$ is the stellar age~\citep[e.g.,][]{Inayoshi+2024}. Since most LRDs remain unresolved, current JWST observations constrain their sizes only to $\lesssim 100~{\rm pc}$, while gravitationally lensed systems improve the limit to $\lesssim 30~{\rm pc}$~\citep[e.g.,][]{Furtak+2023}. Motivated by these constraints, we treat the cluster size $R_{\rm e}$ as a free parameter and consider a range of $5 \lesssim R_{\rm e} \lesssim 100~{\rm pc}$. We assume that the stellar density follows a singular isothermal profile of
\begin{align}
    \rho_{\star}(r) &= \frac{\tilde{\sigma}^{2}}{2 \pi G r^{2}},
\end{align}
where $\tilde{\sigma}$ is the mean stellar velocity dispersion. By normalizing the profile with the total stellar mass, we obtain
\begin{align}
    \tilde{\sigma} \simeq 1.5 \times 10^{3}~{\rm km}~{\rm s}^{-1}~
    \left(\frac{{\cal M}_{\star}}{10^{9} M_{\odot}}\right)^{1/2}
    \left(\frac{R_{\rm e}}{{\rm pc}}\right)^{-1/2},
\end{align}
and thus the density profile is given by
\begin{align}
    \label{eq:stellar_density}
    \rho_{\star} &\simeq 5.9 \times 10^{-15}~{\rm g}~{\rm cm}^{-3} \nonumber \\
    &\quad\times
    \left(\frac{{\cal M}_{\star}}{10^{9} M_{\odot}}\right)
    \left(\frac{R_{\rm e}}{{\rm pc}}\right)^{-1}
    \left(\frac{r}{{\rm pc}}\right)^{-2}.
\end{align}
Inside the stellar cluster, the nuclear SMBH powering the LRD dominates the stellar dynamics within the influence radius
\begin{align}
    r_{h} &= \frac{G M_{\bullet}}{\tilde{\sigma}^{2}} \nonumber \\
    & \simeq 2.0 \times 10^{-3}~{\rm pc} \nonumber \\
    &\quad\times
    \left(\frac{M_{\bullet}}{10^{6} M_{\odot}}\right)
    \left(\frac{{\cal M}_{\star}}{10^{9} M_{\odot}}\right)^{-1}
    \left(\frac{R_{\rm e}}{{\rm pc}}\right).
\end{align}

In the stellar cluster, two-body relaxation redistributes stellar angular momenta and allows 
some stars to fall onto the nucleus. The characteristic timescale for this process is expressed as~\citep[e.g.,][]{Merritt2013}
\begin{align}
    \label{eq:relaxation_time}
    t_{\rm rel} = k \frac{\sigma_{\star}^{3}}{G^{2} \rho_{\star} \bar{m}_{\star} \ln{\Lambda}},
\end{align}
where $\sigma_{\star}$ and $\rho_{\star}$ are the stellar velocity dispersion and density, respectively, and $\bar{m}_{\star} = \langle m_{\star}^{2} \rangle / \langle m_{\star} \rangle$ is the second moment of the stellar mass distribution. Here, $k = 0.34$, $\langle m_{\star}\rangle$ is the mean stellar mass, and $\Lambda = M_{\bullet}/\langle m_{\star}\rangle$. The stellar velocity dispersion is given by $\sigma_{\star}(r) = (G M_{\bullet} / (2r))^{1/2}$. 

To compute the mean and second moment of the stellar mass, we assume a Salpeter mass function, ${\rm d} n_{\star} / {\rm d} m_{\star} \propto m_{\star}^{-\alpha}$ with $\alpha = 2.35$. The minimum stellar mass is set to $m_{\star,{\rm min}} = 0.1~M_{\odot}$. The maximum mass depends on the cluster age. Since the relaxation time decreases with increasing stellar mass, heavier stars are preferentially scattered toward the nucleus. However, they also evolve more rapidly and may die before reaching the central region. The stellar lifetime is approximated by
\begin{align}
    t_{\rm life}(m_{\star}) \simeq 10^{10}~{\rm yr}~
    \left(\frac{m_{\star}}{M_{\odot}}\right)^{-2.5}.
\end{align}
Therefore, the maximum stellar mass is numerically computed by solving
\begin{equation}
t_{\rm rel}(\bar{m}_{\star}(m_{\star,\rm max})) = t_{\rm life}(m_{\star,\rm max}),
\end{equation}
where the relaxation timescale is evaluated at $r=r_{h}$.

\begin{figure}[t]
  \begin{center}
  \includegraphics[keepaspectratio, scale=0.38]{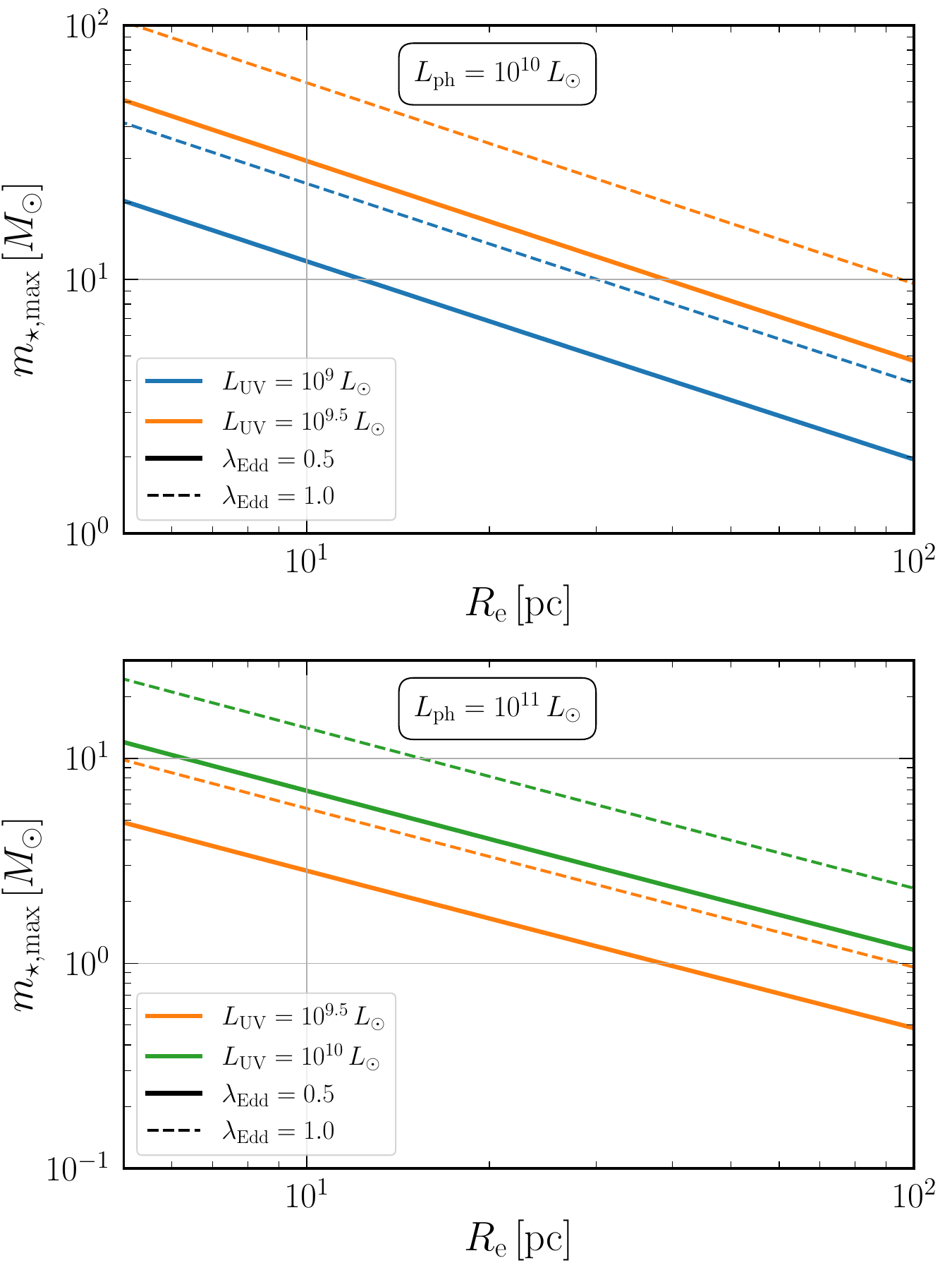}
  \end{center}
  \caption{The maximum stellar mass that can be injected into the envelope within stellar lifetimes as a function of the stellar cluster size, for the cases of $L_{\rm ph} = 10^{10}~L_{\odot}$ (\textit{upper panel}) and $L_{\rm ph} = 10^{11}~L_{\odot}$ (\textit{lower panel}). Different colors show different UV luminosities of the LRD: $L_{\rm UV} = 10^{9}~L_{\odot}$ (blue), $10^{9.5}~L_{\odot}$ (orange), and $10^{10}~L_{\odot}$ (green). The solid and dashed lines correspond to Eddington ratios of $\lambda_{\rm Edd} = 0.5$ and $1.0$, respectively. }
  \label{fig:maximum_mass}
\end{figure}

Figure~\ref{fig:maximum_mass} presents the maximum mass of stars in the cluster that can collide with the envelope surrounding an LRD with luminosities of $L_{\rm ph}=10^{10}~L_{\odot}$ (top) and $10^{11}~L_{\odot}$ (bottom). As the cluster becomes more compact, more massive stars participate in star–envelope collisions because higher stellar densities shorten the relaxation time through more efficient two-body scattering. The maximum stellar mass also decreases with increasing photospheric luminosity. This trend arises because more luminous LRDs are expected to host heavier SMBHs, whose longer relaxation times make it more difficult for massive stars to reach the envelope before the end of their lifetimes. Within the parameter range considered here, the maximum stellar mass exceeds $\sim 3~M_{\odot}$ for clusters with sizes of $\sim 10~{\rm pc}$. Therefore, sufficiently compact clusters can supply evolved massive stars, such as red supergiants with sizes of $R_{\star} \simeq 10^{2}$-$10^{3}~R_{\odot}$, to the envelope, producing observable star–envelope collision events.

The event rate of star-envelope collisions is determined by how frequently stars are scattered into the loss cone, whose angular size is set by the target scale of the interaction. In the present case, the relevant target is the envelope photosphere itself with a loss-cone size of
\begin{align}
    \theta_{\rm lc} \sim \left(\frac{r_{\rm ph}}{r_{h}}\right)^{1/2}.
\end{align}
This differs from standard TDE problems, where the loss-cone size is instead determined by the tidal radius, implying that the collision rate can exceed the TDE rate because the envelope photosphere is typically much larger than the tidal radius. Following \citet{Wang&Merritt2004}, the differential event rate is calculated as 
\begin{align}
    \frac{{\rm d} \dot{N}_{\rm SEC}}{{\rm d} {\cal N}_{\star}} 
    &= \frac{1}{t_{\rm rel} \ln{(2/\theta_{\rm lc})}}.
\end{align}
Using ${\rm d}{\cal N}_{\star} = 4 \pi r^{2} {\rm d}r (\rho_{\star}/\langle m_{\star} \rangle)$, the total event rate becomes
\begin{align}
    \dot{N}_{\rm SEC} &= 4 \pi \int_{0}^{r_{h}} 
    \frac{\ln{\Lambda_{\rm lc} G^{2} \rho_{\star}^{2} r^{3}}}{k \sigma_{\star}^{3}}
    \frac{\bar{m}_{\star}}{\langle m_{\star} \rangle} 
    \frac{{\rm d}r}{r} \nonumber \\
    &= \frac{4 \sqrt{2} \ln{\Lambda_{\rm lc}}}{k \pi} 
    \frac{\tilde{\sigma}^{3}}{G M_{\bullet}} 
    \frac{\bar{m}_{\star}}{\langle m_{\star} \rangle},
\end{align}
where $\ln{\Lambda_{\rm lc}} = \ln{\Lambda}/\ln{(2/\theta_{\rm lc})}$. 

\begin{figure}[t]
  \begin{center}
  \includegraphics[keepaspectratio, scale=0.38]{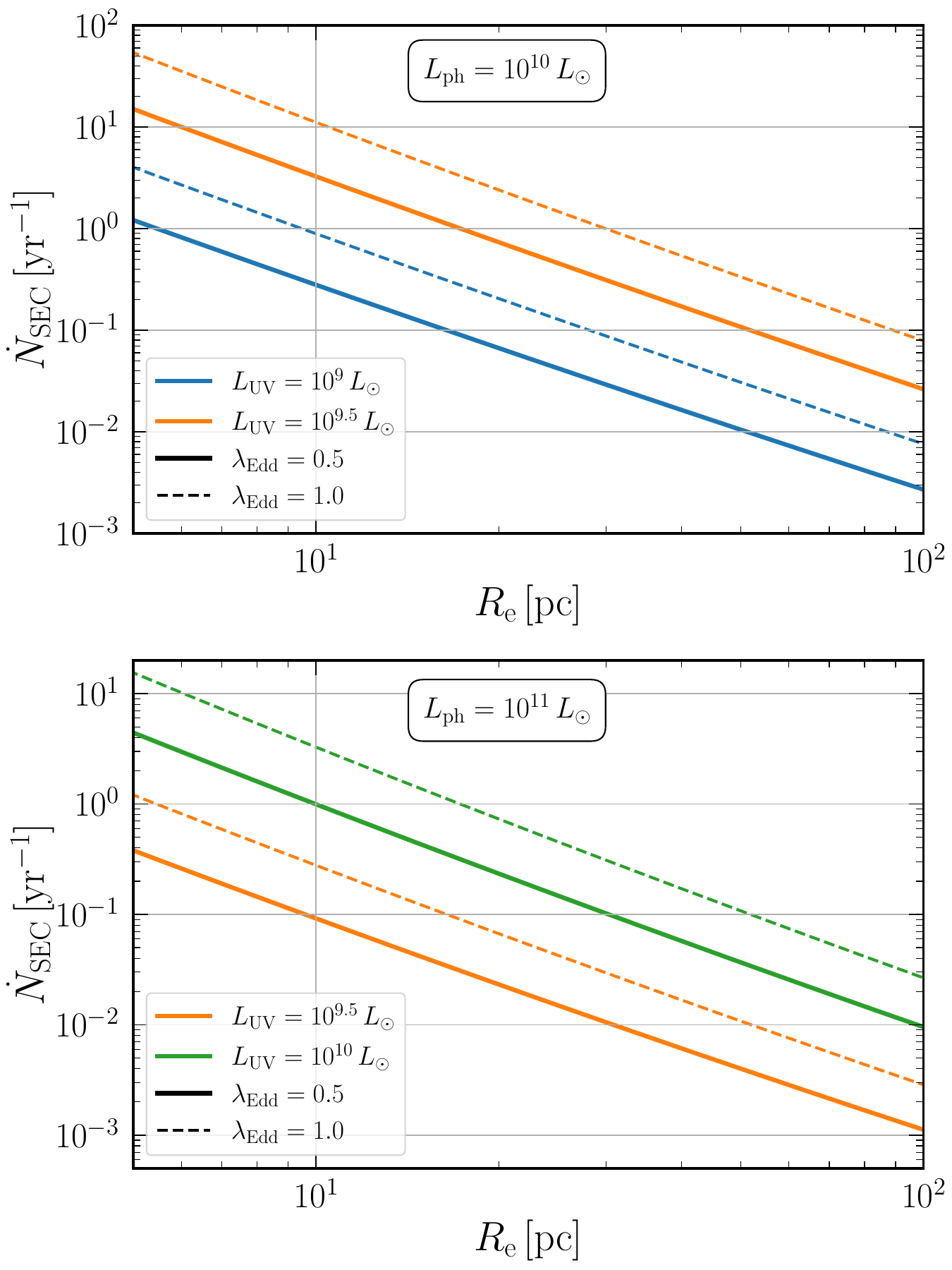}
  \end{center}
  \caption{Event rates of star-envelope collisions. Colors and line styles are the same as in Figure~\ref{fig:maximum_mass}. }
  \label{fig:rate}
\end{figure}

Figure~\ref{fig:rate} shows the event rate of star-envelope collisions as a function of cluster size. 
More compact clusters produce higher event rates because their shorter relaxation times scatter stars toward the SMBH more efficiently. For a fixed UV luminosity (i.e., fixed total stellar mass in the cluster), optically brighter LRDs powered by heavier SMBHs lengthen the relaxation time and reduce the event rate. For cluster sizes of $\sim 10~{\rm pc}$, the predicted event rates reach $\sim 0.3$-$10~{\rm yr}^{-1}$ per galaxy, which is substantially higher than the canonical TDE rate of $\sim 10^{-5}$-$10^{-4}~{\rm yr}^{-1}$~\citep[e.g.,][]{Gezari2021}. Therefore, wide-field observational campaigns monitoring LRDs with sufficient depth and cadence may detect star-envelope collisions as relatively frequent transient events. 

It should be noted that the high event rate is led by the upper end of the stellar mass distribution through $\bar{m}_{\star}$. Although the total event rate is correctly estimated, the rate of collisions involving evolved massive stars ($R_\star \simeq 10^{2-3}~R_{\odot}$ and $m_{\star} \gtrsim 10~M_{\odot}$; see Figure~\ref{fig:maximum_mass}) is likely lower than the value quoted above.

\section{Observability} \label{sec:observability}
As discussed in Section~\ref{subsec:parameter_spectrum}, the peak wavelength of star-envelope collisions lies at $0.2$-$0.5~\mu{\rm m}$ in the rest frame when a red supergiant collides with the envelope. For LRDs at high redshift, the peak wavelength is redshifted to longer wavelengths in the observer frame, shifting the emission into the optical to near-infrared bands.

In addition, as shown in Figures~\ref{fig:obs_lowlum}, the typical duration of these events is $t_{\rm dur} \sim 10^{2.5-3} \, {\rm hr}$~($\simeq 13$-$40~{\rm days}$) for $R_{\star} = 10^{3}~R_{\odot}$ and $M_{\rm env} = M_{\bullet}$. Such relatively short durations make these transients challenging to detect with narrow-field facilities such as the JWST, requiring instead wide-field time-domain surveys. Given that the emission falls in the optical to near-infrared range, the Vera C. Rubin Observatory Legacy Survey of Space and Time (LSST) and the Nancy Grace Roman Space Telescope (RST) are well suited for detecting these events.

\begin{figure*}[t]
  \begin{center}
  \includegraphics[keepaspectratio, scale=0.4]{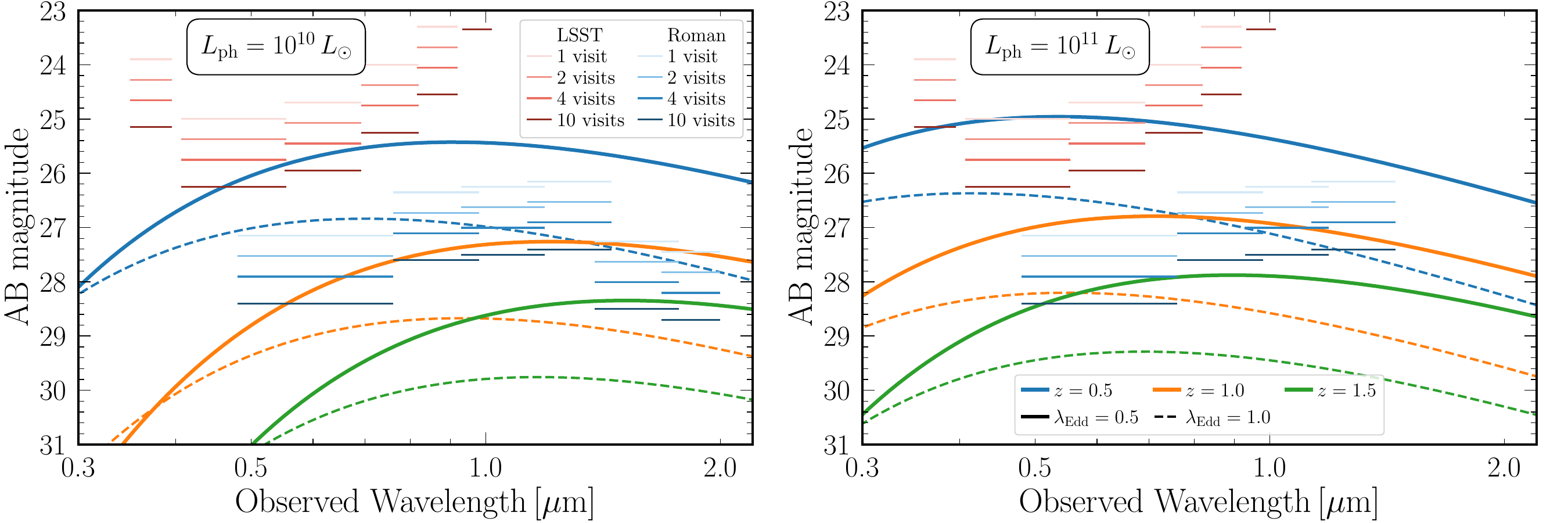}
  \end{center}
  \caption{SEDs of star-envelope collisions in the observer frame for the cases of $L_{\rm ph} = 10^{10}~L_{\odot}$ (\textit{left panel}) and $L_{\rm ph} = 10^{11}~L_{\odot}$ (\textit{right panel}). The horizontal axis represents the AB magnitude. Different colors show sources at different redshifts (blue: $z = 0.5$, orange: $z = 1.0$, and green: $z = 1.5$). The solid and dashed lines correspond to Eddington ratios of $\lambda_{\rm Edd} = 0.5$ and $1.0$, respectively. The sensitivities of RST and LSST, summarized in Table~\ref{table:Roman_sensitivity} and Table~\ref{table:LSST_sensitivity}, respectively, are also plotted assuming ${\rm SNR} = 5$. Considering the duration of the transient (see text), the sensitivity curves are shown for $1, 2, 4$, and $10$ visits. The stellar radius colliding with the envelope and the envelope mass are fixed at $R_{\star} = 10^{3}~R_{\odot}$ and $M_{\rm env} = M_{\bullet}$, respectively. }
  \label{fig:obs_spec}
\end{figure*}

Figure~\ref{fig:obs_spec} shows the SEDs of star-envelope collisions in the observer frame for the case of a red supergiant colliding with an envelope whose mass is comparable to that of the SMBH ($R_{\star} = 10^{3}~R_{\odot}$ and $M_{\rm env} = M_{\bullet}$). Considering single-exposure observations, LSST is unable to detect these transients for sources at $z \gtrsim 0.5$. In contrast, RST can detect them over the redshift range $0.5 \lesssim z \lesssim 1.0$. For $z \sim 0.5$, multi-band detections with RST are possible. For example, in the case of $L_{\rm ph} = 10^{10}~L_{\odot}$, the peak AB magnitude is $\sim 25.5$-$27~{\rm mag}$ with peak wavelengths of $\sim 0.7$-$0.8~\mu{\rm m}$ in the observer's frame, allowing detection even in a single RST exposure. 

\begin{table}
    \caption{Single exposure Sensitivities of RST. The signal to noise ratio is set to SNR~$=10$~\citep{Rose+2021}. }
    \label{table:Roman_sensitivity}
    \centering
    \small
    \setlength{\tabcolsep}{15pt}
    \begin{tabular}{ccc}
        \hline
        Filter
        & Wavelength $[\mu{\rm m}]$ 
        & Sensitivity \\
        \hline\hline
        F062 
        & $0.48$-$0.76$ 
        & $26.4$ \\
        F087    
        & $0.76$-$0.977$ 
        & $25.6$ \\
        F106
        & $0.927$-$1.192$ 
        & $25.5$ \\
        F129
        & $1.131$-$1.454$ 
        & $25.4$ \\
        F158
        & $1.380$-$1.774$ 
        & $26.5$ \\
        F184
        & $1.683$-$2.000$ 
        & $26.7$ \\
        \hline
    \end{tabular}
\end{table}

\begin{table}
    \caption{Single exposure sensitivities of LSST. The signal to noise ratio is set to SNR~$=5$~\citep{Bianco+2022}. }
    \label{table:LSST_sensitivity}
    \centering
    \small
    \setlength{\tabcolsep}{15pt}
    \begin{tabular}{ccc}
        \hline
        Filter
        & Wavelength $[\mu{\rm m}]$ 
        & Sensitivity \\
        \hline\hline
        $u$ 
        & $0.3724 \pm 0.0463/2$ 
        & $23.0$ \\
        $g$    
        & $0.4807 \pm 0.1485/2$ 
        & $25.0$ \\
        $r$
        & $0.6221 \pm 0.1399/2$ 
        & $24.7$ \\
        $i$
        & $0.7559 \pm 0.1286/2$ 
        & $24.0$ \\
        $z$
        & $0.8680 \pm 0.1040/2$ 
        & $23.3$ \\
        $y$
        & $0.9753 \pm 0.0862/2$ 
        & $22.1$ \\
        \hline
    \end{tabular}
\end{table}

In addition, in the case of $L_{\rm ph} = 10^{10}~L_{\odot}$ and $z \sim 0.5$, the relatively long duration in the observer frame enhances detectability. Given a rest-frame duration of $t_{\rm dur,rest} \sim 10^{2.5-3}~{\rm hrs}$, the observed duration becomes $t_{\rm dur,obs} \sim t_{\rm dur,rest} \times (1 + z) \sim 20$-$60~{\rm days}$. With typical survey cadences of $\sim 5~{\rm days}$ for both RST~\citep{Rose+2021} and LSST~\citep{Bianco+2022}, these events can be monitored over $\sim 4$-$12$ visits. Such repeated observations improve detectability, enabling LSST to also identify these events despite its lower single-exposure sensitivity. Additionally, the observed duration increases with redshift, which partially compensates for the decreasing flux and aids in the detection of more distant sources. Overall, RST is expected to detect these events up to $z \sim 1.0$, with similar trends holding for the case of $L_{\rm ph} = 10^{11}~L_{\odot}$. 

At higher redshifts ($z \gtrsim 1.0$), detection becomes increasingly difficult. This can be attributed to two main effects. First, the flux decreases and falls below the sensitivity limits of RST and LSST. Second, the peak wavelength shifts to longer wavelengths, moving out of the RST bandpass. 

Conversely, detection is also challenging at very low redshifts ($z < 0.5$), despite the higher flux. In this regime, the peak wavelength shifts toward the blue and approaches the edge of the RST band coverage. For instance, for $L_{\rm ph} = 10^{10}~L_{\odot}$ at $z \simeq 0$, the peak wavelength is $\sim 0.5~\mu{\rm m}$, which lies near the blue edge of the RST filters, making it difficult to robustly determine the location of the spectral peak. In such cases, complementary observations with the $u$- and $g$-bands of LSST would be important, provided that accurate cross-calibration is achieved.

\section{Discussions} \label{sec:discussion}

\subsection{Low-redshift LRD Selection by Wide-Field Surveys} \label{subsec:LRD_selection}
Flares associated with star–envelope collisions in LRDs are expected to be detectable at relatively low redshifts ($0.5 \lesssim z \lesssim 1.5$; see Section~\ref{sec:observability}). Prior identification of host LRDs within a survey field would therefore facilitate the discovery of these short-duration transients. LRDs are characterized by distinctive V-shaped SEDs in the rest-frame UV–optical bands~\citep[e.g.,][]{Greene+2024,Kocevski+2025,Hviding+2025} with an inflection point near the Balmer limit ($\lambda_{\rm rest}=3645$~\AA)~\citep[e.g.,][]{Setton+2025}. At $z \simeq 0.5$-$1.5$, this spectral feature is redshifted to $\lambda_{\rm obs} \simeq 0.5$-$0.9~\mu{\rm m}$, which is covered by the filter sets of both RST and LSST.

\begin{figure}[t]
  \begin{center}
  \includegraphics[keepaspectratio, scale=0.41]{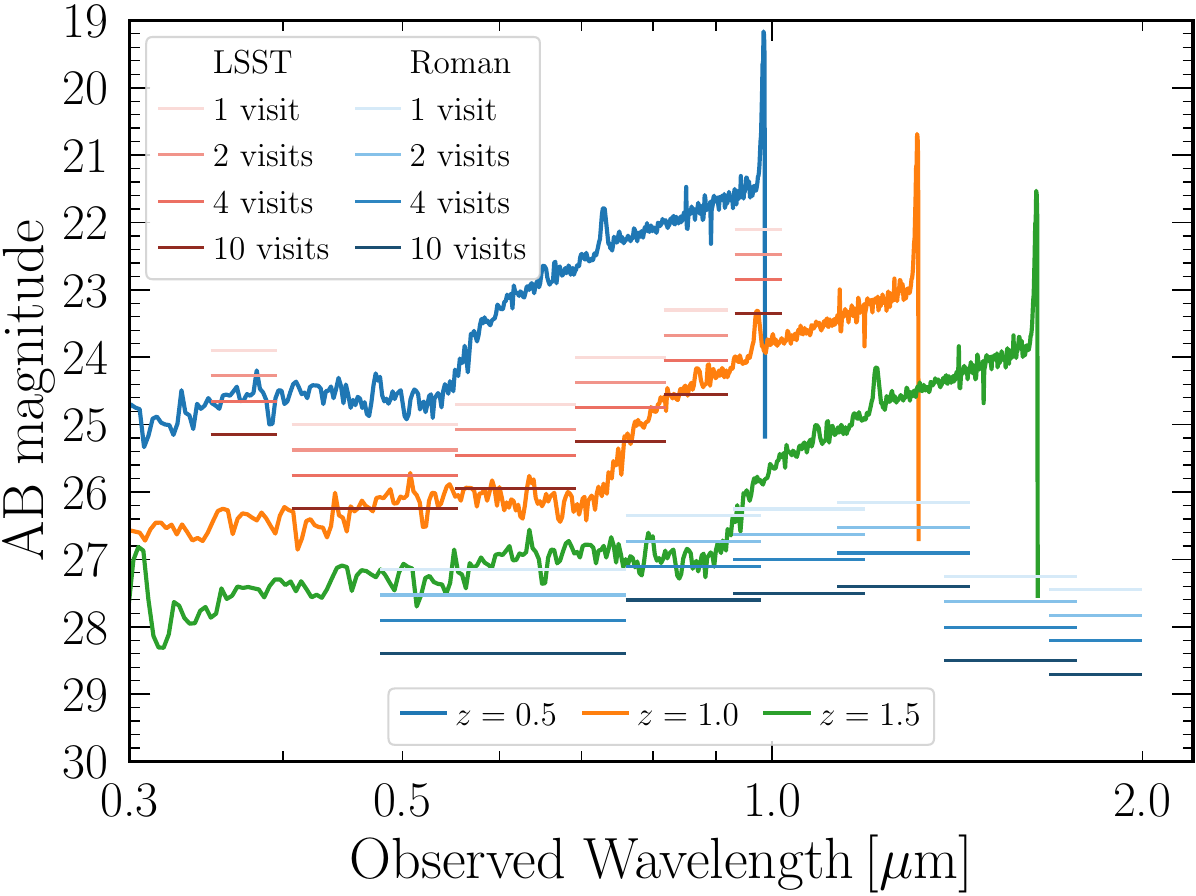}
  \end{center}
  \caption{Examples of LRD SEDs at different redshifts (blue: $z = 0.5$, orange: $z = 1.0$, and green: $z = 1.5$) in the observer frame. The horizontal axis represents the AB magnitude. Each SED is constructed from the spectrum of A2744-QSO1 ($z = 7.045$; \citealt{Furtak+2024}) by accounting for redshift effects. The sensitivities of RST and LSST are also plotted, as in Figure~\ref{fig:obs_spec}. }
  \label{fig:LRD_spec}
\end{figure}

To examine the detectability of low-redshift LRDs with RST and LSST, we adopt the NIRSpec/PRISM spectrum of A2744-QSO1 at $z = 7.045$ as a representative LRD spectrum~\citep{Furtak+2024} and rescale its lensing-corrected flux density to lower redshifts, as shown in Figure~\ref{fig:LRD_spec}. The characteristic V-shaped SED can be well captured by the RST filter set for sources at $1.0 \lesssim z \lesssim 1.5$ in a single or two visits. For sources at $z \sim 0.5$, however, the inflection point shifts close to the blue edge of the shortest-wavelength RST filter, making it difficult for RST alone to robustly identify the V-shaped spectral feature. In this regime, the $u$- and $g$-bands of LSST provide important complementary coverage. Figure~\ref{fig:LRD_spec} further suggests that, for sources at $z \lesssim 0.5$, LSST alone may already be capable of identifying the V-shaped SED even in a single visit. Therefore, combining RST and LSST provides a powerful strategy for efficiently discovering low-redshift LRDs.

\subsection{Cosmic Event Rates of Star-Envelope Collisions in LRD Nuclei} \label{subsec:detection_number}
To estimate the total number of detectable star–envelope collisions, both the event rate per LRD (Section~\ref{sec:event_rate}) and the cosmic abundance of LRDs over the relevant redshift range ($0.5 \lesssim z \lesssim 1.5$; Section~\ref{sec:observability}) must be considered. Thus far, low-redshift LRD searches suggest that the LRD number density peaks at $z\simeq 4$-$6$ and declines rapidly toward lower redshifts \citep[e.g.,][]{Kocevski+2025,Ma+2026,Inayoshi2025}. Current constraints imply a plausible range of $10^{-10} \lesssim n_{\rm LRD} \lesssim 10^{-6}~{\rm cMpc}^{-3}$ over $0.5 \lesssim z \lesssim 1.5$~\citep{Ma+2026,Lin+2026a,Park+2026,Lin+2026b}.

For the planned wide-tier RST survey covering $\sim 20~{\rm deg}^{2}$ \citep{Rose+2021}, the expected number of LRDs is
\begin{align}
    N_{\rm LRD,RST} \sim 5.5~\left(\frac{n_{\rm LRD}}{10^{-7} \, {\rm cMpc}^{-3}}\right),
\end{align}
based on a comoving volume of $\sim 5.5 \times 10^{7}~{\rm cMpc}^{3}$ over this redshift range. Here, $n_{\rm LRD}$ denotes the LRD number density integrated over luminosity. According to \citet{Ma+2026}, the lower limit of the rest-frame $5500$~\AA~luminosity is $\sim 10^{43}~{\rm erg}~{\rm s}^{-1}$, and such sources are detectable in a single RST exposure (see Section~\ref{subsec:LRD_selection}). Assuming $L_{\rm ph} = 10^{10}~L_{\odot}$, the total number of detectable star-envelope collisions over the planned RST survey is 
\begin{align}
    N_{\rm SEC} &\simeq \dot{N}_{\rm SEC} N_{\rm LRD,RST} T_{\rm RST} \nonumber \\
    &\sim 3.3 - 110 \, \left(\frac{n_{\rm LRD}}{10^{-7} \, {\rm cMpc}^{-3}}\right),
\end{align}
where we adopt $\dot{N}_{\rm SEC} \sim 0.3$-$10~{\rm yr}^{-1}$ (see Section~\ref{sec:event_rate}) and take $T_{\rm RST} \sim 2~{\rm yr}$ as the planned mission lifetime of RST~\citep{Rose+2021}. These results suggest that if the cosmic abundance of LRDs at $0.5 \lesssim z \lesssim 1.5$ exceeds $\sim 3 \times 10^{-8}~{\rm cMpc}^{-3}$, the RST may detect at least one star-envelope collisions during its mission lifetime. The LRD abundance at $z < 0.45$ is estimated as $\sim 1.6 \times 10^{-9}~{\rm cMpc}^{-3}$~\citep{Park+2026}, implying that the detection of star-envelope collision events may be difficult. However, interpolating the observed abundance between the low-redshift regime at $z < 0.45$ and intermediate regimes at $2 < z < 4$ \citep{Ma+2026}, the LRD number density would be comparable to or even higher than $10^{-8}~{\rm cMpc}^{-3}$.

\subsection{Model Constraints from (Non-)Detection} \label{subsec:observational_implication}
The detection of star-envelope collisions would provide important insights into the nature of LRDs. Most importantly, such events require the presence of a dense gaseous envelope surrounding the SMBH. Therefore, detecting these transients in LRDs would provide direct evidence for the envelope scenario.

These events can also constrain the physical properties of the envelope itself. If the spectral peak of such a transient is detected, both the peak luminosity and temperature can be measured. As discussed in Section~\ref{sec:collision}, these observables depend on the envelope luminosity $L_{\rm ph}$, temperature $T_{\rm ph}$, and mass $M_{\rm env}$. Among them, $L_{\rm ph}$ and $T_{\rm ph}$ can already be constrained from the optical spectrum of the host LRD itself. Star-envelope collisions therefore provide a unique probe of the envelope mass, which is otherwise difficult to determine from the LRD spectra alone. In addition, the duration and luminosity of the transient depend on the radius of the colliding star, potentially providing information on the stellar population in the surrounding cluster. Moreover, if multiple events are detected, the size of the stellar cluster may also be constrained by comparing the observed event rate with theoretical predictions.

On the other hand, star-envelope collisions may remain undetected even in long-term surveys with LSST and RST because our estimates rely on several simplifying assumptions. For example, if the surrounding stellar cluster is less compact, the event rate can be substantially lower (Section~\ref{sec:event_rate}). In addition, the transients themselves may be too faint to detect even when the rate is sufficiently high. Such faintness can arise from either a lower envelope mass or a smaller stellar radius. A less massive envelope leads to a lower gas density near the envelope surface, reducing the ejecta mass produced during the collision and consequently lowering the peak luminosity. Similarly, a smaller stellar radius reduces the interaction cross section with the envelope, resulting in a smaller ejecta mass and hence dimmer emission.

\subsection{Emergence of TDE Emission from the Gaseous Envelope} \label{subsec:TDE_implication}
If stars are scattered onto sufficiently low-angular-momentum orbits that enter the loss cone, whose size is determined by the tidal radius, they will eventually be tidally disrupted by the central SMBH. In addition, stars may lose angular momentum through interactions with the envelope gas, further enhancing the probability of tidal disruption. Such events can produce additional transient signals in LRDs~\citep[e.g.,][]{Inayoshi+2024}. However, if a sufficiently massive envelope surrounds the SMBH, the observable signatures of TDEs may be obscured or even suppressed. 

We examine the conditions under which TDE emission can escape from the envelope. A star with mass $M_{\star,{\rm TDE}}$ and radius $R_{\star,{\rm TDE}}$ is tidally disrupted when the pericenter distance $r_{\rm p}$ is smaller than the tidal radius, 
\begin{align}
    r_{\rm T} &= \left(\frac{M_{\bullet}}{M_{\star,{\rm TDE}}}\right)^{1/3} R_{\star,{\rm TDE}} \nonumber \\
    &\simeq 7.0 \times 10^{12}~{\rm cm}~
    \frac{M_{\bullet,6}^{1/3} r_{\star,{\rm TDE}}}{m_{\star,{\rm TDE}}^{1/3}},
\end{align}
where $r_{\star,{\rm TDE}} = R_{\star,{\rm TDE}}/R_{\odot}$ and $m_{\star,{\rm TDE}} = M_{\star,{\rm TDE}}/M_{\odot}$~\citep{Rees1988}. After disruption, approximately half of the debris becomes unbound, while the remaining half remains gravitationally bound to the SMBH~\citep[e.g.,][]{Rees1988,Evans&Kochanek1989}. The bound debris initially follows highly eccentric orbits and subsequently loses orbital energy through stream-stream collisions, where earlier fallback material, which is more tightly bound and returns faster, intersects with later fallback material on longer fallback timescales, producing shocks typically near the apocenter of the orbits~\citep[e.g.,][]{Rees1988,Shiokawa+2015,Ryu+2023b,Price+2024,Steinberg&Stone2024}. This interaction drives rapid circularization and the formation of an accretion disk around the SMBH.

The most tightly bound debris initially has a specific binding energy of 
\begin{align}
    \epsilon_{\rm debris} &\simeq -\left(\frac{G M_{\bullet}}{r_{\rm p}}\right)
    \left(\frac{R_{\star,{\rm TDE}}}{r_{\rm p}}\right) \nonumber \\
    &\simeq -1.9 \times 10^{17}~{\rm erg}~{\rm g}^{-1} \, 
    \frac{\beta^{2} m_{\star,{\rm TDE}}^{2/3} M_{\bullet,6}^{1/3}}{r_{\star,{\rm TDE}}},
\end{align}
where $\beta = r_{\rm T}/r_{\rm p}$ is the penetration factor. After complete circularization and disk formation, the specific energy becomes 
\begin{align}
    \epsilon_{\rm circ} &= -\frac{G M_{\bullet}}{2 r_{\rm p}} \nonumber \\
    &\simeq -9.6 \times 10^{18}~{\rm erg}~{\rm g}^{-1}~
    \frac{\beta m_{\star,{\rm TDE}}^{1/3} M_{\bullet,6}^{2/3}}{r_{\star,{\rm TDE}}},
\end{align}
Since $|\epsilon_{\rm circ}| \gg |\epsilon_{\rm debris}|$, a substantial amount of energy is released during circularization. The total energy dissipated by the bound debris can be estimated as
\begin{align}
    E_{\rm circ} &\simeq |\epsilon_{\rm circ}|\left(\frac{M_{\star,{\rm TDE}}}{2}\right) \nonumber \\
    &\simeq 9.5 \times 10^{51}~{\rm erg}~
    \frac{M_{\bullet,6}^{2/3} m_{\star,{\rm TDE}}^{4/3} \beta}{r_{\star,{\rm TDE}}}.
\end{align}
If the energy released during circularization exceeds the binding energy of the envelope defined by
\begin{align}
    E_{\rm env} &= \frac{G M_{\rm env} M_{\bullet}}{r_{\rm ph}} \nonumber \\
    &\simeq 2.7 \times 10^{49}~{\rm erg}~
    \left(\frac{M_{\bullet}}{10^{6} \, M_{\odot}}\right)
    \left(\frac{M_{\rm env}}{M_{\odot}}\right)
    \left(\frac{r_{\rm ph}}{10^{16} \, {\rm cm}}\right)^{-1},
\end{align}
the envelope can be significantly perturbed or partially unbound, allowing the TDE emission to escape. From this condition, we obtain the critical envelope mass,
\begin{align}
    M_{\rm env,crit} \simeq 1.8 \times 10^{2}~M_{\odot}~
    \frac{\beta (r_{\rm ph}/10^{16} \, {\rm cm}) m_{\star,{\rm TDE}}^{4/3}}{M_{\bullet,6}^{1/3} r_{\star,{\rm TDE}}},
\end{align}
above which TDE signals are expected to be strongly suppressed by the presence of the envelope. Conversely, the detection of TDE-like emission from LRDs would imply an envelope mass below $100~M_{\odot}$. 

Finally, we note that for intermediate envelope masses, TDE emission may be reprocessed by the envelope and emerge with modified spectral properties. A detailed prediction of such reprocessed signals is beyond the scope of this paper.

\section{Conclusions}
We have investigated the observational signatures of star-envelope collisions based on a scenario in which LRDs consist of an envelope surrounded by a dense stellar cluster. When a star collides supersonically with the envelope surface, a strong shock forms and drives the expansion of shocked material. The ejecta is initially opaque, but becomes transparent as it expands. Once the photon diffusion timescale becomes comparable to the dynamical timescale, radiation can escape efficiently and produce an observable transient signal. Based on this framework, we modeled the emission and estimated the characteristic duration, peak luminosity, and peak temperature of the resulting flares. Our main findings are summarized as follows: 

\begin{enumerate}
\item Both the duration and luminosity of the transient increase with the radius of the colliding stars because larger ejecta masses lead to longer diffusion times and greater radiated energies. The most promising events are collisions between red supergiants and massive envelopes surrounding SMBHs. For fiducial parameters ($L_{\rm ph} = 10^{10}~L_{\odot}$ and $T_{\rm ph} = 6000~{\rm K}$), we find durations of $t_{\rm dur} \simeq 10^{2.5}$-$10^{3}~{\rm hrs}$, peak luminosities of $L_{\rm peak} \simeq 10^{41.5}$-$10^{42}~{\rm erg}~{\rm s}^{-1}$, and rest-frame peak wavelengths of $\lambda_{\rm peak} \simeq 0.3$-$0.5~\mu{\rm m}$. 

\item Maximum masses of stars that participate in star-envelope collisions are determined by the competition between stellar relaxation and stellar evolution: stars must migrate toward the SMBH before exhausting their lifetimes. More compact stellar clusters shorten the relaxation times and thus allow more massive stars to contribute. For clusters with characteristic sizes of $\sim 10~{\rm pc}$, stars with masses of $\sim 10$-$100~M_\odot$ can reach the envelope within their lifetimes. These compact environments also enhance the collision rate itself. For fiducial parameters, we obtain event rates of $\gtrsim 0.3~{\rm yr}^{-1}$ per LRD host galaxy, substantially higher than canonical TDE rates.

\item The predicted transients are potentially detectable with future wide-field surveys such as RST and LSST, particularly for sources at $0.5\lesssim z\lesssim1$. At higher redshifts, the detectability rapidly decreases because the peak luminosity falls below the survey sensitivity and the emission shifts outside the observable bands. Although low-redshift LRDs are expected to be rare, current constraints on their abundance still allow at least of order unity sources within the planned RST survey volume. Combined with the relatively high event rate per galaxy, this implies that multiple star-envelope collisions could be discovered during the mission lifetime.

\item If a star undergoing a stellar-envelope collides is subsequently tidally disrupted, the resulting TDE emission may remain obscured by the envelope. However, when the envelope mass is as small as $\lesssim 100~M_{\odot}$, the TDE energy can exceed the envelope binding energy and disrupt the envelope itself, allowing the TDE emission to emerge.
\end{enumerate}

Detection of star-envelope collisions through wide-field surveys with RST and LSST would provide direct evidence for the envelope scenario of LRDs and offer a unique probe of the physical properties of the envelope and the surrounding stellar cluster. In particular, combining transient observations with the spectra of host LRDs may constrain quantities such as the envelope mass and the stellar population, which are otherwise difficult to infer directly. While the present study adopts a simplified emission model, more realistic predictions will require future radiation-hydrodynamical simulations of star-envelope interactions.

\begin{ack}
We thank Koutarou Kyutoku, Kohta Murase, and Xuejian Shen for motivating this work. We also thank Takashi Hosokawa and Kunihito Ioka for fruitful discussions and comments. 
\end{ack}

\section*{Funding}
T.S. is supported by JST SPRING grant No. JPMJSP2110, JSPS KAKENHI grant No. JP26KJ0041. K.I. acknowledges support from the National Natural Science Foundation of China (12573015, W2532003), the Beijing Natural Science Foundation (IS25003), and the China Manned Space Program (CMS-CSST-2025-A09).


\bibliographystyle{apj}
\bibliography{ref_NT,ref_LRD}

\appendix 
\section{Dependence on the Photospheric Luminosity} \label{appsec:parameter}
\begin{figure}[t]
  \begin{center}
  \includegraphics[keepaspectratio, scale=0.35]{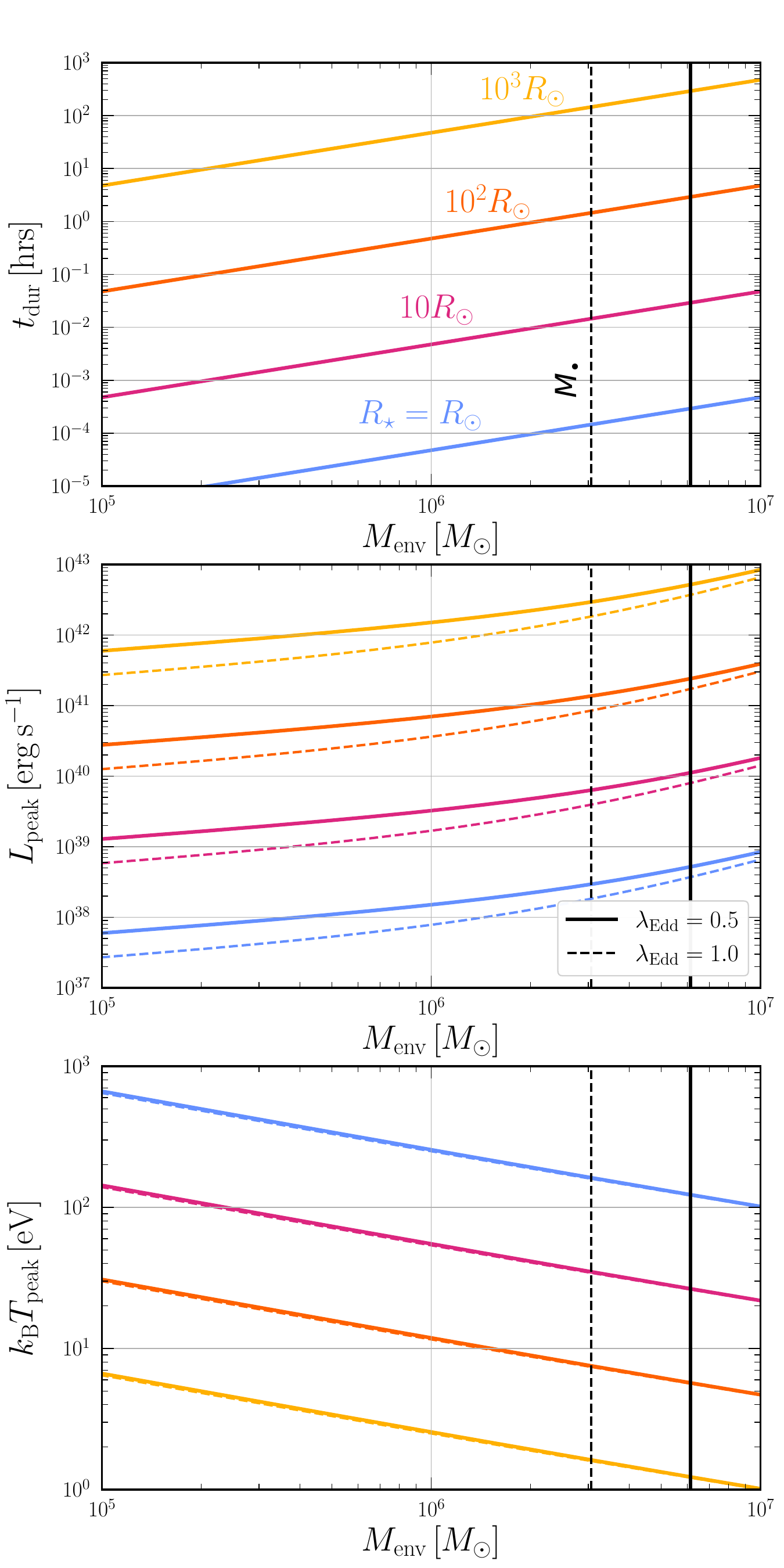}
  \end{center}
  \caption{Same as Fig.~\ref{fig:obs_lowlum}, but with the photospheric luminosity set to $L_{\rm ph} = 10^{11}~L_{\odot}$. }
  \label{fig:obs_highlum}
\end{figure}

Figure~\ref{fig:obs_highlum} shows the observable properties of star-envelope collisions for $L_{\rm ph} = 10^{11}~L_{\odot}$. The qualitative trends are similar to those for $L_{\rm ph} = 10^{10}~L_{\odot}$ shown in Figure~\ref{fig:obs_lowlum}. However, for fixed envelope mass and stellar radius, the duration becomes shorter, while the peak luminosity and peak temperature become higher. This behavior mainly arises because a larger photospheric luminosity corresponds to a larger photospheric radius and hence a lower envelope density at the photosphere. The resulting ejecta mass is therefore smaller, leading to a shorter diffusion timescale and a higher ejecta temperature. 

\end{document}